\documentclass[prb,twocolumn,showpacs,groupedaddress]{revtex4}%
\usepackage{amsmath}
\usepackage{graphicx}
\usepackage{amsfonts}
\usepackage{amssymb}%
\begin{document}
\title{Artifact of the phonon-induced localization by variational calculations in the spin-boson
model}
\author{Zhi-De Chen}\email[Author to whom correspondence should be addressed: ]{tzhidech@jnu.edu.cn}
\author{Hang Wong}
\affiliation{Department of Physics, Jinan University, Guangzhou
510632, China}
%\date{July 24, 2002}arXiv:0705.1670

\begin{abstract}
We present energy and free energy analyses on all variational
schemes used in the spin-boson model at both $T=0$ and $T\not=0$. It
is found that all the variational schemes have fail points, at where
the variational schemes fail to provide a lower energy (or a lower
free energy at $T\not=0$) than the displaced-oscillator ground state
and therefore the variational ground state becomes unstable, which
results in a transition from a variational ground state to a
displaced oscillator ground state when the fail point is reached.
Such transitions are always misidentified as crossover from a
delocalized to localized phases in variational calculations, leading
to an artifact of phonon-induced localization. Physics origin of the
fail points and explanations for different transition behaviors with
different spectral functions are found by studying the fail points
of the variational schemes in the single mode case.

\end{abstract}
\pacs{03.65.Yz, 03.65.Xp, 73.40.Gk} \maketitle
\section{Introduction}
Variational method is one of the important tools in tracing the
ground state of many body systems.\cite{gp} In dealing with
systems having interaction with phonons, the variational method
based on displaced-oscillator state (DOS) plays an important role.
Some well-known examples in condensed matter physics are
polaron-phonon, exciton-phonon systems,\cite{llp,la,ss} and
spin-boson model, which is an important toy model in dissipative
quantum systems.\cite{leg,weiss,ss} In fact, the coupling to a
phonon bath in all the mentioned systems has the same bilinear
coupling form. One important issue in the mentioned systems is the
so-called phonon-induced localization. In exciton-phonon (or
polaron-phonon) systems, the phonon-induced localization is called
the self-trapped transition, while in the spin-boson model, it is
stated as the crossover from a delocalized to localized
phases.\cite{leg,weiss,sp,keh}  Such a crossover is now considered
as some kind of quantum phase transition called boundary phase
transition.\cite{qpt,qp} In general, variational calculation has
been serving as an important tool in dealing with the
phonon-induced localization, however, the predicted transitions in
exciton-phonon (or polaron-phonon) systems turned out to be false
ones but the answer to why the variational calculation fails was
not found.\cite{gl} Variational calculation on the crossover
problem in the spin-boson model has a long history and the first variational scheme is based on the DOS.\cite{leg,wz,sh,nie}%,cy,cyw,dj,ct,wh}% but the result is far from satisfactory.
It was shown that, in strong coupling regime, the coupling to phonon
bath can lead to both displacement and deformation of the oscillator
and the so-called displaced-squeezed-state (DSS) is energy favorite,
leading to a variational calculation based on the DSS.\cite{cy,cyw}
Later, a hybrid variational scheme based on the DSS was also
developed.\cite{dj} At finite temperature regime, the variational
calculation based on the DOS was done for both the Ohmic and
sub-Ohmic cases and a discontinuous crossover was
found,\cite{sh,ct,wh,cc} and recently, it is shown that the
variational calculation based on the DSS fails in thermodynamical
limit.\cite{c1}

Despite of its long history, the result from variational
calculations is far from satisfactory. The basic problem is that the
predicted crossover point is variational-scheme dependent. For
example, in the Ohmic case, the crossover point by the variational
calculation based on the DOS was $\alpha_c=1$,\cite{wz} while that
based on the DSS gave $\alpha_c=2,$\cite{cy,cyw} and the hybrid
variational scheme predicted a $\alpha_c$ lay between.\cite{dj} The
transition behavior predicted by the variational calculations is
also in controversy. The variational calculations based on both the
DOS and DSS predicted a continuous transition in the Ohmic case but
discontinuous transition behavior was also found in the hybrid
variational scheme.\cite{dj} In the sub-Ohmic case, all the
variational calculations predicted a discontinuous
transition,\cite{ct,wh} but non-perturbative numerical calculations,
like numerical renormalization group (NRG) and density-matrix
renormalization group (DMRG), found a continuous transition
instead.\cite{bulla,b1,wh1} The situation in the case of $T\not=0$
is also confusing. Variational calculations predicted a
discontinuous transition at $T\not=0$ even in the super-Ohmic
case,\cite{ch} a result which is in confliction with the known
conclusion that the crossover only happens in the Ohmic and
sub-Ohmic cases.\cite{leg,weiss,sp,keh} All these discrepancies make
the crossover predicted by the variational calculations in question,
e.g., is it an artifact as in the case of polaron-phonon systems?
Clearing up this issue is important for understanding the result by
variational calculations and further application of variational
methods to other systems. This is the motivation of the present
paper.

By employing energy analysis on the variational schemes used in
the spin-boson model, we shown that the predicted crossover is
just an artifact of the variational schemes. We will show that all
the variational schemes mentioned above have fail points, at where
variational schemes fail to provide a lower energy than that of
the displaced-oscillator ground state, leading to a transition
from the variational ground state to the displaced-oscillator
ground state when the fail point is reached. In variational
calculations, such a transition is always misidentified as a
crossover from a delocalized to localized phases and the fail
point is misinterpreted as a crossover point. The result we found
can help to resolve the controversy mentioned above.

The rest of the paper is organized as follows: The fail points of
the variational schemes at both $T=0$ and $T\not=0$ are demonstrated
in the next section. In Sec III, physical origin of the fail point
and different transition behaviors in both the Ohmic and sub-Ohmic
cases are analyzed by studying the variational schemes in the single
mode case. Conclusion and discussion are presented in the last
section.

\section{Fail points of the variational schemes used in the spin-boson model}
The Hamiltonian of the spin-boson model is given by (setting
$\hbar=1$)\cite{leg,weiss}
\begin{equation}
\hat{H}=-\frac{\Delta_0}{2}\hat{\sigma}_x+\sum_k
\hat{b}_k^{\dagger}\hat{b}_k\omega_k+\hat{\sigma}_z\sum_k c_k(\hat{b}_k^{\dagger}+\hat{b}_k),
\end{equation} where $\hat{\sigma}_i$ $(i=x,y,z)$ is the Pauli matrix,
$\hat{b}_k$ $(\hat{b}_k^{\dagger})$ is the annihilation (creation)
operator of the $k$th phonon mode with energy $\omega_k$ and $c_k$
is the corresponding coupling parameter. As is known, the solution
of this model depends on the so-called bath spectral function that
is defined as $ J(\omega)=\pi
\sum_kc_k^2\delta(\omega-\omega_k).\cite{leg,weiss} $ Usually the
spectral function has a power law form, i.e.,
\begin{equation}
J(\omega)=\frac{\pi}{2}\alpha \omega^s\omega_D^{1-s},~~0<\omega<\omega_D,
\end{equation} %and for the cases of both Ohmic and sub-Ohmic dissipation,
where $\omega_D$ is the cut-off frequency and $\alpha$ is a
dimensionless coupling strength which characterizes the
dissipation strength.\cite{leg,weiss} %The super-Ohmic case is not considered here since no crossover was found in that case.
The property of the bath is characterized by the parameter $s$, i.e., $0<s<1$, $s=1$, and $s>1$ represent respectively the sub-Ohmic,
Ohmic, and super-Ohmic cases.

It is well known that, with the bilinear coupling form given in the last term of Eq.(1), the Hamiltonian can be %in adiabatic approximation,
diagonalized by the so-called displaced-oscillator transformation in the case of $\Delta_0=0$. The ground state in this case %of the model  in Eq.(1)
is the displaced-oscillator ground state\cite{leg,weiss,mah}
\begin{equation}
|\Phi_0\rangle=\frac{1}{\sqrt{2}}(|\phi_+\rangle|\uparrow\rangle+|\phi_-\rangle|\downarrow\rangle),
\end{equation} where $ |\phi_{\pm}\rangle=\exp\{\mp \sum_k
(c_k/\omega_k)~(\hat{b}_k-\hat{b}_k^{\dagger})\}|0\rangle, $ is the
DOS, while $|\uparrow\rangle$ and $|\downarrow\rangle$ are the
eigenstates of $\hat{\sigma}_z$. All the variational schemes used in
the spin-boson model start from the DOS; analysis in details will be
presented in the following.
\subsection{Fail points of the variational schemes at $T=0$}
As we have mentioned in the Sec I, there are three kinds of
variational schemes used in the spin-boson model at $T=0$ and the
key difference is the trial ground state. The trial ground state for
the variational calculation based on the DOS is chosen as\cite{wz}
\begin{equation}
|\Psi_0\rangle=\frac{1}{\sqrt{2}}(|\psi_+\rangle|\uparrow\rangle+|\psi_-\rangle|\downarrow\rangle),
\end{equation}
where
\begin{equation} |\psi_{\pm}\rangle%=U|0\rangle
=\exp\left\{\mp \sum_k
g_k~(\hat{b}_k-\hat{b}_k^{\dagger})\right\}|0\rangle,
\end{equation}
and $g_k$ is the variational parameter. The trial ground state based
on the DSS is given by\cite{cy,cyw}
\begin{equation}
|\tilde{\Psi}_0\rangle=\frac{1}{\sqrt{2}}(|\varphi_+\rangle|\uparrow\rangle+|\varphi_-\rangle|\downarrow\rangle),
\end{equation}
where
\begin{equation}
|\varphi_{\pm}\rangle=\exp\{\mp
\sum_k\frac{c_k}{\omega_k}~(\hat{b}_k-\hat{b}_k^{\dagger})\}\exp\{-\sum_k\gamma_k(\hat{b}_k^2-\hat{b}_k^{\dagger
2})\}|0\rangle,
\end{equation}
is the DSS and $\gamma_k$ is the variational parameter. The third
variational scheme is a hybrid scheme of the above two variational
schemes and the trial ground state is\cite{dj}
\begin{equation}
|\tilde{\Phi}_0\rangle=\frac{1}{\sqrt{2}}(|\tilde{\varphi}_+\rangle|\uparrow\rangle+|\tilde{\varphi}_-\rangle|\downarrow\rangle),
\end{equation}
with
\begin{equation}
|\tilde{\varphi}_{\pm}\rangle=\exp\{\mp
\sum_kg_k(\hat{b}_k-\hat{b}_k^{\dagger})\}\exp\{-\sum_k\gamma_k(\hat{b}_k^2-\hat{b}_k^{\dagger
2})\}|0\rangle,
\end{equation}
here both $g_k$ and $\gamma_k$ are the variational parameters.

All the variational schemes have almost the same calculation process
and the main routine is as follows. Firstly, one used the trial
ground state $|\Phi_{tgs}\rangle$ to calculate the trial ground
state energy $E_t=\langle \Phi_{tgs}|\hat{H}|\Phi_{tgs}\rangle$,
then the following variational condition
\begin{equation}
\delta E_t/\delta g_k=\delta E_t/\delta \gamma_k=0,
\end{equation}
can help to determine the variational parameters. In what follows,
the trial ground state  with the variational parameters determined
from the above variational condition is called the variational
ground state $|\Psi_v\rangle$. Variation of the tunneling splitting
of the variational ground state with the dissipative strength
$\alpha$ can tell how the crossover happens. Since all three
variational calculations give the same crossover behavior in both
the Ohmic and sub-Ohmic cases, here we shall use the variational
calculation based on the DOS as an example to show the details.

For variational scheme based on the DOS, the variational condition
(10) leads to
\begin{equation}
g_k=\frac{c_k}{\omega_k+ \Delta_0\exp\{-2\sum_kg_k^2\}},
\end{equation}
from which the self-consistent equation of the dressing factor $K_1$ can be found\cite{wz,nie,wh}
\begin{equation}
K_1=f(K_1)\equiv
\exp\left\{-\frac{2}{\pi}\int_0^{\omega_D}\frac{J(\omega)}{(\omega+
K_1\Delta_0)^2}d\omega\right\},
\end{equation}
here $K_1=\Delta/\Delta_0$ and $\Delta$ is the tunneling splitting
in presence of dissipation. The trivial and non-zero solutions of
Eq.\ (12) represent respectively the localized ($\Delta=0$) and
delocalized ($\Delta\not=0$) phases. Variational calculation
determines the crossover by examining the evolution of solutions
of Eq.\ (12) with the dissipation strength $\alpha$. It turned out
that, in the Ohmic case, there is only one non-zero solution which
decreases to zero continuously as $\alpha$ approaches to 1 in the
limit of $\Delta_0/\omega_D\ll1$, showing a continuous crossover
happens at $\alpha_c=1$. The situation in the sub-Ohmic case is
more complicated; there are two non-zero solutions ($K_1'$ and
$K_1''$) which disappear abruptly when $\alpha$ reaches some
critical value. It has been shown that the crossover in this case
is discontinuous and the crossover point should be determined
according to the Landau theory.\cite{ct,wh} The situation for
other variational schemes is the same except for the values of the
critical points. It should be noted that the discontinuous
transition behavior in the Ohmic case found in the hybrid
variational scheme\cite{dj} can also be found in other variational
schemes and the key point is to increase the value of
$\Delta_0/\omega_D$. We find that, for $\Delta_0/\omega_D\ge 0.5$,
a discontinuous transition appears in the Ohmic case for the
variational scheme based on the DOS. The change of transition
behaviors is shown in Fig.\ 1. Explanation for this result will be
presented in next section.
\begin{figure}[h]
\centering
\includegraphics[width=0.48\textwidth]{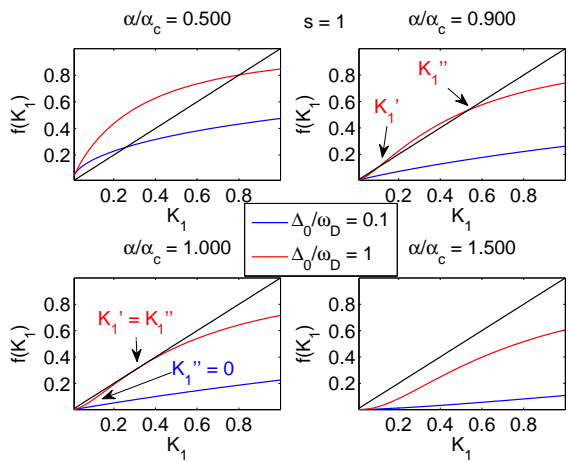}
\caption{(Color online) The evolution of solutions of the
self-consistent equation derived from the variational calculation
based on the DOS in the Ohmic case. In the case of
$\Delta_0/\omega_D=0.1$, there is only one non-zero solution and the
transition behavior is continuous. However, in the case of
$\Delta_0/\omega_D=1$, there are two non-zero solutions $K_1'$ and
$K_1''$ and the transition behavior becomes discontinuous as
$\alpha$ approaches $\alpha_c$. }
\end{figure}

Now we turn to show how variational schemes fail as $\alpha$ increases. %However, we shall show that energy analysis tells a different story.
As we have mentioned, the starting point of all the variational
schemes is DOS. Consequently, for a successful variational scheme,
the variational ground states $|\Psi_v\rangle$ should be a better
approximation to the true ground state than the displaced-oscillator
ground state $|\Phi_0\rangle$ , which implies that $|\Psi_v\rangle$
should have a lower energy than $|\Phi_0\rangle$, e.g., the
following condition
\begin{equation}
E_v=\langle \Psi_v|\hat{H}|\Psi_v\rangle< E_0=\langle
\Phi_0|\hat{H}|\Phi_0\rangle,
\end{equation}
should hold in a variational calculation, otherwise {\it the
variational scheme fails}. Surprisingly, it is found that all the
variational schemes fail when $\alpha$ increases to some critical
value in both the Ohmic and sub-Ohmic cases. In the sub-Ohmic case,
the variational ground state has a lower energy than the
displaced-oscillator ground state when $\alpha\ll \alpha_c$. As
$\alpha$ increases, both $E_v$ and $E_0$ decreases but $E_0$
decreases in a way faster than $E_v$, we have $E_v=E_0$ when
$\alpha=\alpha_c$, and $E_v>E_0$ as $\alpha>\alpha_c$, showing that
the variational scheme fails at $\alpha=\alpha_c$. Typical results
for the variational calculation based on the DOS in the case of
$s=0.5$ and $\Delta_0/\omega_D=0.1$ are shown in Fig.\ 2. For the
Ohmic case with $\Delta_0/\omega_D\ll 1$, the energy difference
$E_v-E_0$ decreases to zero continuously as $\alpha\rightarrow
\alpha_c$. The results of the variational calculation based on the
DOS is shown in Fig.\ 3. However, as $\Delta_0/\omega_D$ increases
to some larger values, the self-consistent equation has two non-zero
solutions as shown in Fig.\ 1, and the situation becomes the same as
in the sub-Ohmic case.

Analytically, the fail point $\alpha_c$ of the variational scheme
can be determined by
\begin{equation}
E_v=E_0.
\end{equation}
In the case of the variational scheme based on the DOS, this leads
to
\begin{equation}
\alpha_c=\frac{\Delta_0(K_1-K_0)}{
\int_0^{\omega_D}(\frac{\omega}{\omega_D})^{s-1}\left(\frac{K_1\Delta_0}
{\omega+K_1\Delta_0}\right)^2d\omega}, \end{equation} where $
K_0=\langle
\phi_+|\phi_-\rangle=\exp\{-\frac{2}{\pi}\int_0^{\omega_D}\frac{J(\omega)}{\omega^2}d\omega\},
$ is the adiabatic dressing factor. In the Ohmic case, we have $s=1$
and $K_0=0$ and the fail point is
\begin{equation}
\alpha_c= 1+K_1(\Delta_0/\omega_D).
\end{equation}
The above equation is just the crossover point predicted by the
variational calculation in the limit of $\Delta_0/\omega_D\ll
1$.\cite{wz,nie}
\begin{figure}[h]
\centering
\includegraphics[width=0.48\textwidth]{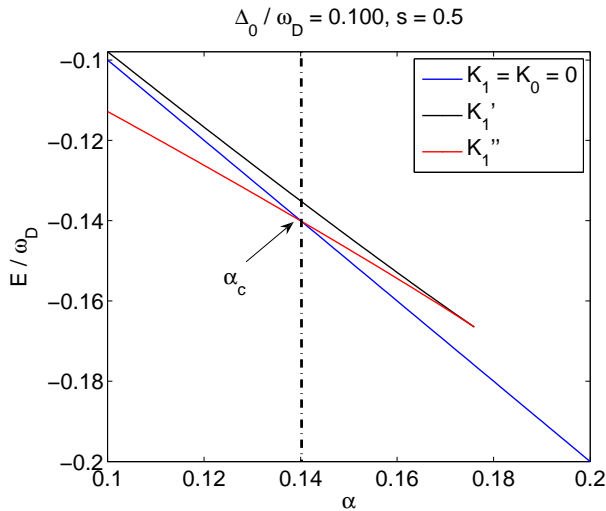}
\caption{(Color online) Variation of both the variational ground
state energy $E_1(K_1'')$ and the displaced oscillator ground state
energy $E_0(K_1=0)$ as a function of $\alpha$ in the sub-Ohmic case
with $s=0.5$ and $\Delta_0/\omega_D=0.1$. $E_1(K_1')$ represents the
energy maximum lies between. For $\alpha>\alpha_c$,
$E_1(K_1'')>E_0(K_1=0)$ demonstrates the failure of the variational
scheme.}
\end{figure}
On the other hand, for the variational scheme based on the DSS, the
fail point equation tells
\begin{equation}
\frac{1}{2}\Delta_0D=\sum_k\sinh^2(2\gamma_k)\omega_k
|_{\alpha=\alpha_c},
\end{equation}
where $\gamma_k$ is determined by $
e^{4\gamma_k}=\sqrt{1+4D\Delta_0c_k^2/\omega_k^3} $ and $D=\langle
\varphi_+|\varphi_-\rangle$ is the dressing factor.\cite{cy,cyw} The
right-hand-side of the above equation represents the energy gain due
to the deformation of the phonon state (the squeezed
effect).\cite{cyw} This energy gain depends on the spectral function
as well as the phonon density of states which is assumed to be
$\rho(\omega)=\omega^{n-1}/\omega_D^n$, where $n$ is the dimension
degree of the bath in the long-wavelength approximation.\cite{cy} In
the Ohmic case ($s=1$), it can be found that
\begin{equation}
\alpha_c\simeq n+1,~~~~{\rm for}~~~\Delta_0/\omega_D\ll 1.
\end{equation}
Once again this result coincides with the crossover point predicted
by the corresponding variational calculation.\cite{cy,note} The
calculation in the hybrid variational scheme is rather complicated
and we find it is hard to find the crossover point analytically in
this case.

%In sub-Ohmic case, it has been shown that the transition is discontinuous, and by Landau theory,\cite{gol,cl} the
%critical point is exactly given by Eq.(5).\cite{wh}
\begin{figure}[h]
\centering
\includegraphics[width=0.48\textwidth]{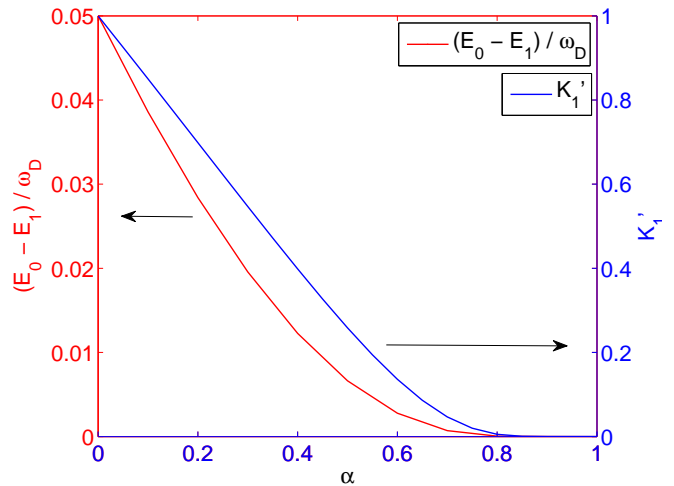}
\caption{(Color online) Variation of both the energy difference
$(E_0-E_1)$ (with scale on the left) and the non-zero solution
$K_1'$ (with scale on the right. Note that this $K_1'$ is the
non-zero solution in the Ohmic case, not one of the solutions in the
sub-Ohmic case mentioned above) of self-consistent equation with the
dissipation strength $\alpha$ in the Ohmic case with
$\Delta_0/\omega_D=0.1$. As $\alpha$ increases, both $(E_0-E_1)$ and
$K_1'$ decrease continuously to zero as $\alpha\rightarrow 1$. }
\end{figure}
The above analysis provides a different explanation for the
crossover predicted by the variational calculations.
%As a conclusion, the crossover predicted by the variational calculation can be described alternatively from the above analysis.
When $\alpha<\alpha_c$, we have $E_v<E_0$ and the variational
ground state is stable. However, for $\alpha>\alpha_c$, the
variational scheme fails to give a lower energy and the
variational ground state becomes unstable, which results in a
transition from the variational ground state to the displaced
oscillator ground state at $\alpha=\alpha_c$. Since the
variational ground state has a non-zero tunneling splitting while
the displaced-oscillator ground state has zero tunneling
splitting, such a transition is misidentified as the crossover
from the delocalized to localized phases, leading to an artifact
of phonon-induced localization. It is important to note that the
transition point from the variational ground states to the
displaced-oscillator ground state is determined by Eq.\ (14). In
the Ohmic case with $\Delta_0/\omega_D\ll1$, the transition is
continuous, while in the sub-Ohmic case or Ohmic case with a large
$\Delta_0/\omega_D$, the transition is discontinuous and in this
case, the transition point is given by Eq.\ (14) according to
Landau theory.\cite{wh} Based on the above result, it is easy to
understand why different variational schemes predicted different
crossover points.
\subsection{Fail point of the variational scheme at $T\not=0$}
In the case of $T\not=0$, there is only one kind of variational
scheme based on the DOS. The application of other variational
schemes mentioned above is in question since the variational
scheme based on the DSS was shown to fail in thermodynamical
limit.\cite{c1} Variational calculation at $T\not=0$ is different
from the case of $T=0$.\cite{sh,ct} Firstly, we perform a
displaced-oscillator transformation to the Hamiltonian given in
Eq.\ (1),
\begin{equation}
\hat{H}'=U^{-1}\hat{H}U=\hat{H}_0+\hat{V},
\end{equation}
where
\begin{equation}
U=\exp\{\sigma_z\sum_kf_k(b_k-b_k^{\dagger}),
\end{equation}
and $f_k$ is the variational parameter. It can be found that
\begin{equation}
\hat{H}_0=\frac{\Delta}{2}\sigma_x+\sum_k(b_k^{\dagger}b_k+\frac{1}{2})\omega_k+\sigma_z^2\sum_k(
\omega_kf_k^2-2f_kc_k),
\end{equation}
\begin{equation}
\hat{V}=V_+\sigma_++V_-\sigma_-+V_0\sigma_z,
\end{equation}
where
\begin{equation}
V_+=V_-^*=\frac{\Delta_0}{2}\exp\{-2\sum_kf_k(b_k-b_k^{\dagger})\}-\frac{\Delta}{2},
\end{equation}
$V_0=\sum_k(b_k^{\dagger}+b_k)(c_k-f_k)$, and $\Delta$ is the
tunneling splitting in the presence of dissipation at $T\not=0$. Up
to the second order approximation, the up-bound of the free energy
can be found by the Bogoliubov-Feynman theorem and the result is
(setting the Boltzmann constant $k_B=1$)\cite{ct}
\begin{equation}
A_B=-T\ln[2\cosh(\beta\Delta/2)]+\sum_k(\omega_kf_k^2-2f_kc_k),
\end{equation}
where $\beta=1/T$ and the variational condition in the present
case reads
\begin{equation}
\delta A_B/\delta f_k=0,
\end{equation}
which leads to
\begin{equation}
f_k=\frac{c_k}{\omega_k+\Delta \coth(\beta\omega_k/2)\tanh(\beta \Delta/2)},
\end{equation}
and a self-consistent equation of the dressing factor $D=\Delta/\Delta_0$ can be found
\begin{equation}
D=\exp\left\{-\alpha\int_0^{\omega_D}\frac{J(\omega)\coth(\beta\omega/2)d\omega}{[\omega+D\Delta_0\coth(\beta\omega/2)\tanh(\beta \Delta/2)]^2}\right\},
\end{equation}
substituting the above result back to Eq.\ (24) and the up-bound of
the free energy for the variational thermodynamical ground state can
be found. In analog to the case at $T=0$, we have a
displaced-oscillator thermodynamical ground state and its up-bound
of free energy can be found by setting $f_k=c_k/\omega_k$, the
result is
\begin{equation}
F_0=-T\ln
2-\frac{\alpha}{2}\int_0^{\omega_c}d\omega\frac{J(\omega)}{\omega}.
\end{equation}
It is clear that the displaced-oscillator ground state corresponds
to the trivial solution of the self-consistent equation (27), e.g.,
$A_B(D=0)=F_0$. By the way, as in the case of $T=0$, we should have
$A_B<F_0$ for a successful variational scheme and the fail point
equation is
\begin{equation}
F_0=A_B.
\end{equation}
\begin{figure}[h]
\centering
\includegraphics[width=0.48\textwidth]{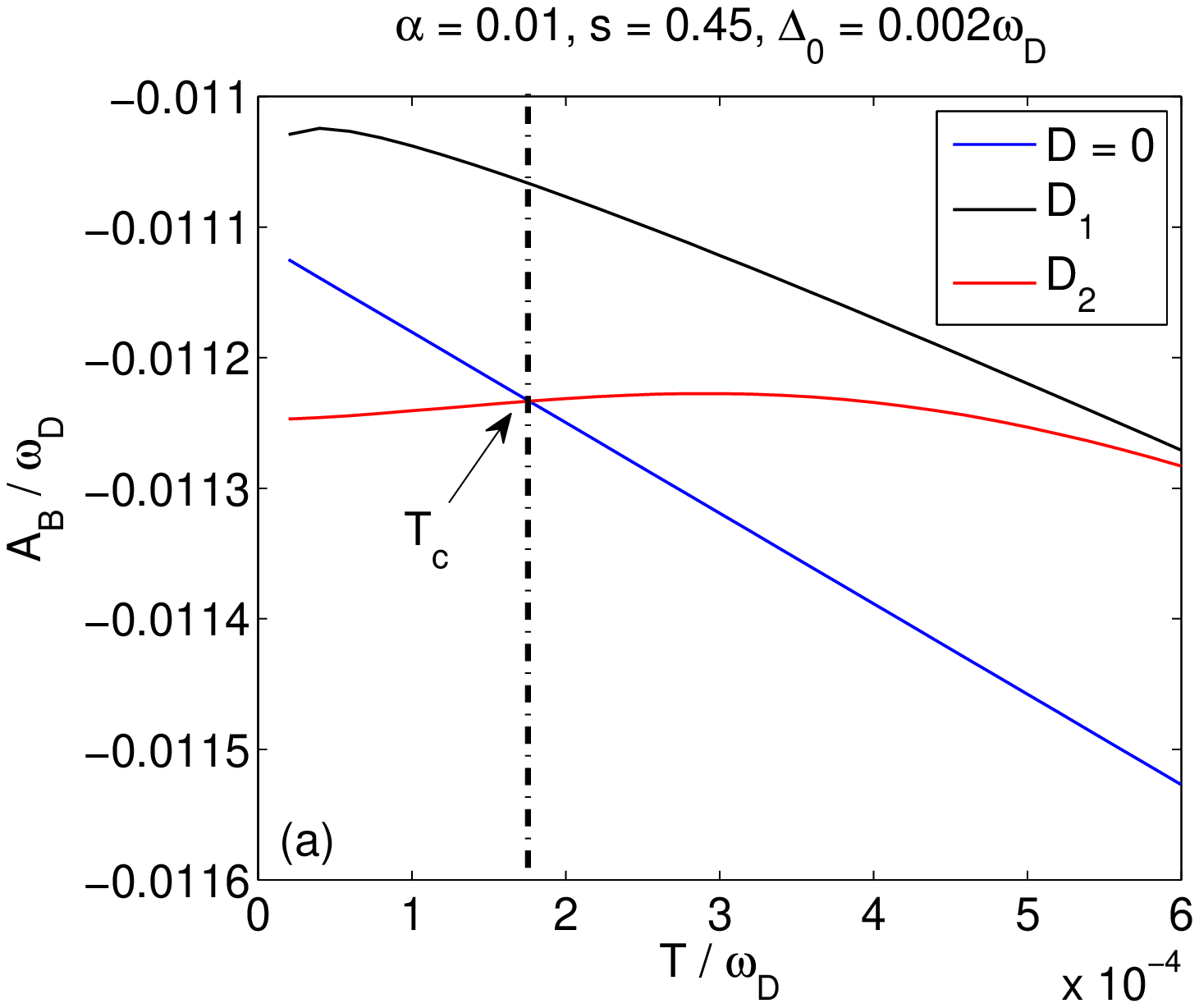}
\includegraphics[width=0.48\textwidth]{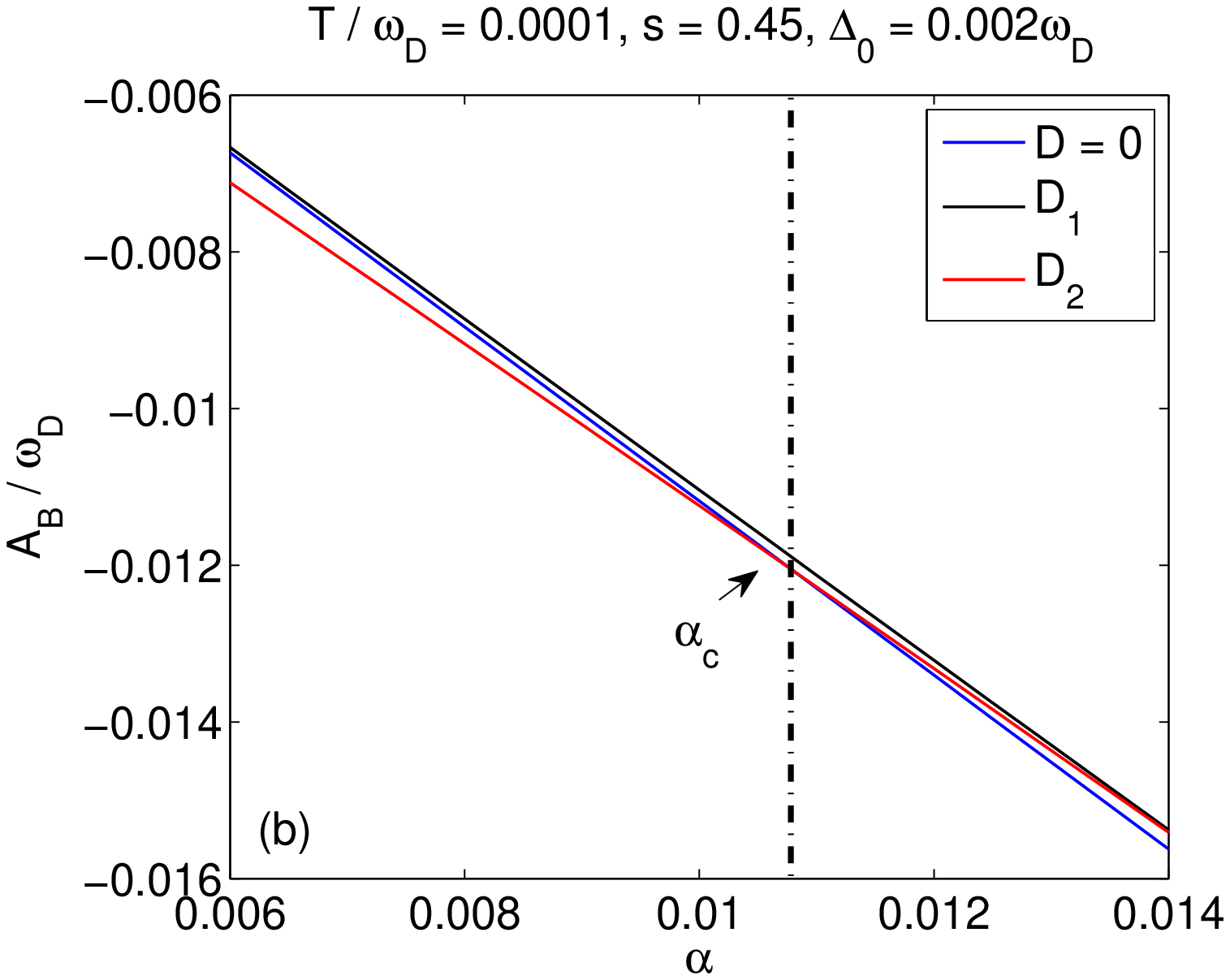}
\caption{(Color online) Comparison of the free energies obtained
from the self-consistent equation in the case of $s=0.45$ and
$\Delta_0/\omega_D=0.002$. The trivial solution $D=0$ represents the
displaced-oscillator thermodynamical ground state. Two non-zero
solutions $D_2$ and $D_1$ represents the variational thermodynamical
ground state and a maximum lies between. (a) Illusion for the
failure of the variational scheme for a fixed $\alpha=0.01$. At the
beginning of the low temperature region, $A_B(D_2)<F_0=A_B(D=0)$ and
the variational thermodynamical ground state is stable, when $T$
increases to some critical point $T_c$ we have $A_B(D_2)=A_B(D=0)$.
For $T>T_c$ we have $A_B(D_2)>A_B(D=0)$, showing the failure of
variational scheme at $T=T_c$. (b) Illusion for the failure of the
variational scheme for a fixed $T/\omega_D=10^{-4}$ in the case of
$\Delta_0/\omega_D=0.002$. The situation is the same as in (a),
variational scheme fails when $\alpha>\alpha_c$.}
\end{figure}
Numerical analysis shows that, for a fixed temperature, the
variational scheme fails when $\alpha$ increases to some critical
point $\alpha_c$, while for a fixed $\alpha$, the variational
scheme fails when $T$ increases to some critical point $T_c$.
Typical results are shown in Fig.\ 4. As a matter of fact, the
situation is more or less that same as in the case of $T=0$. The
variational thermodynamical ground state becomes unstable when the
fail point is reached and a transition from the variational
thermodynamical ground state to the displaced-oscillator
thermodynamical ground state happens, which is misidentified as
the crossover from the delocalized to localized phases. It is
worth noting that such a transition happens for $0<s\le 2$ in the
case of $T\not=0$.\cite{ch} Based on the above analysis, the
transition for $1<s\le 2$ (i.e., in the super-Ohmic case) is not a
true delocalized-localized phase transition, hence it is not in
confliction with the known result.
\section{Further analysis on fail points of the variational schemes}
In this section, we shall try to answer two questions: why the
variational schemes fail and why we have different transition
behaviors (continuous and discontinuous) when the fail point is
reached?  To this goal, we turn to study the oversimplified
spin-boson model, i.e., a two-level system coupled to just one
phonon mode, the Hamiltonian in the single phonon mode case is given
by\cite{ss,ir}
\begin{equation}
\hat{H}_1=-\frac{1}{2}\Delta_0\hat{\sigma}_x+\omega
\hat{b}^{\dagger}\hat{b}+\lambda\hat{\sigma}_z (\hat{b}^{\dagger}+\hat{b}).
\end{equation}
The reason why we study this simple model is as follows. Firstly,
the fail point analysis presented above can be used to this model to
see whether fail point exists in the single mode case. Secondly, the
ground state of this model can be known analytically in an
approximate way or more exactly by the numerical
diagoanlization.\cite{ss,ir} The comparison between the variational
ground state with fail point and the true ground state can help to
give the answer.
\subsection{Fail point in the single mode case}
In the case of $T=0$, the fail point analysis for multi-mode case
can be directly generalized to the present case. For example, the
fail point equation for this model can be directly deduced from the
corresponding equations given before. For clarity, we shall use the
variational scheme based on the DOS as an example to show the
details. From Eq.\ (15), one can find that the fail point equation
in the present case is given by
\begin{equation}
\frac{1}{2}(k_1-k_0)\Delta_0=(\lambda_c/\omega)^2\frac{k_1^2(\Delta_0/\omega)}{(1+k_1\Delta_0/\omega)^2},
\end{equation}
where $k_1$ is given by the self-consistent equation
\begin{equation}
k_1=\exp\left\{-\frac{2\lambda^2}{(\omega+k_1\Delta_0)^2}\right\},
\end{equation}
and $k_0=e^{-2(\lambda/\omega)^2}$. The above equation, in general,
have three solutions $k_0*,k_1'$, and $k_1''$ where
$k_0*<k_1'<k_1''$. Numerical analysis shows there exists a frequency
threshold $\omega_c$, variational schemes fail only for
$\omega<\omega_c$ at $\lambda=\lambda_c$, which is
$\omega$-dependent.
%The fail point curves for both variational schemes are shown in Fig.3.
It turns out that, for $\omega>\omega_c$, the variational ground
state is always stable, i.e., $E(k_1)<E(k_0)$, and the tunneling
splitting $k_1\Delta_0$ decreases continuously with increasing
$\lambda$. However, for $\omega<\omega_c$, the tunneling splitting
shows a discontinuous jump as $\lambda$ approaching $\lambda_c$, at
where the variational ground state becomes unstable. Figure 5 shows
the variation of both the energy and tunneling splitting around the
fail point. In fact, such a discontinuous jump was firstly reported
in Ref.\ \onlinecite{ss} and stated it as an artifact of the
variational calculation. Our analysis shows clearly this artifact is
due to the fail point of the variational scheme. Nevertheless, the
discontinuous jump in the single mode case has some important
difference from the discontinuous transition in the multi-mode case.
First of all, the discontinuous jump does not exactly lead to a
crossover from the delocalized to localized phases since both phases
have non-zero tunneling splitting. The most important difference is
that the transition is not from the variational ground state to the
displaced-oscillator ground state, but to a state closed to it. This
is because $k_0$ is not a solution of the self-consistent equation
(32), a situation that is qualitatively different from the
multi-mode case where $K_0=0$ is the trivial solution of
self-consistent equation (12) in both the Ohmic and sub-Ohmic cases.
Since the transition determined from the self-consistent equation
can only happens between the solutions of this equation, the
transition cannot be from the variational ground state to the
displaced-oscillator ground state directly, nevertheless the
difference becomes very small when $\omega/\Delta_0\le 1/5$.
\begin{figure}[h]
\centering
\includegraphics[width=0.48\textwidth]{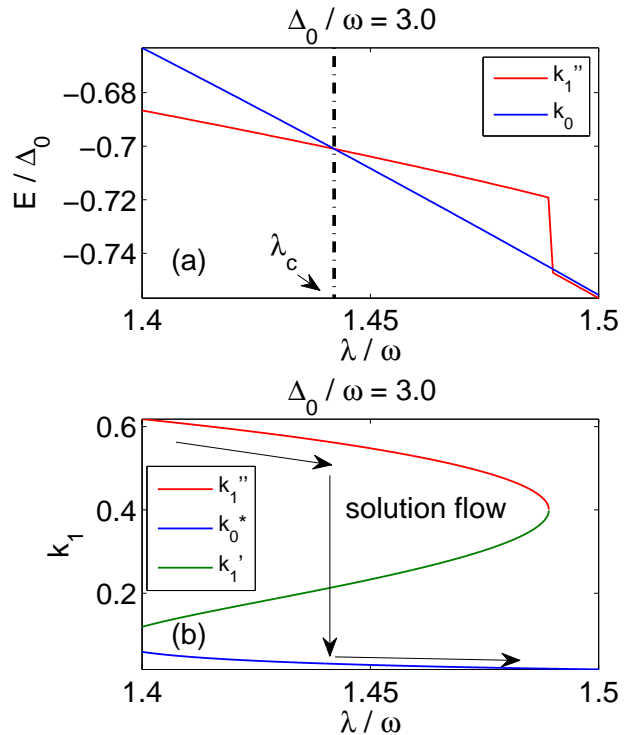}
\caption{(Color online) Illusion of the fail point of the
variational scheme based on the DOS in the single-mode case with
$\Delta_0/\omega=3.0$. In the present case, the self-consistent
equation has three solutions, $k_1''$, $k_0^*$, and $k_1'$, which
represent respectively the variational ground state, a state closed
to displaced-oscillator ground state, and the energy maximum lies
between. (a) Variation of the two ground state energies with
$\lambda$. The variational ground state energy becomes larger than
the displaced-oscillator ground state when $\lambda>\lambda_c$,
showing the failure of the variational scheme. (b) Solutions of the
self-consistent equation as a function of $\lambda$. When the fail
point is reached, the variational ground state become unstable and
tunneling splitting will have discontinuous jump, i.e., from
$k_1''\Delta_0$ to $k_0^*\Delta_0$. }
\end{figure}

Fail points of other variational schemes can be analyzed in the same
way. The situation is almost the same for all three variational
schemes except the values of $\omega_c$ and the corresponding
$\lambda_c$. Figure 6 shows the boundary lines of the fail points as
a function of $\Delta_0/\omega$ and $\lambda/\omega$ for all the
three variational schemes. As one can see from Fig.\ 6, all
variational schemes fail in the low frequency and strong coupling
regions.
\begin{figure}[h]
\centering
\includegraphics[width=0.48\textwidth]{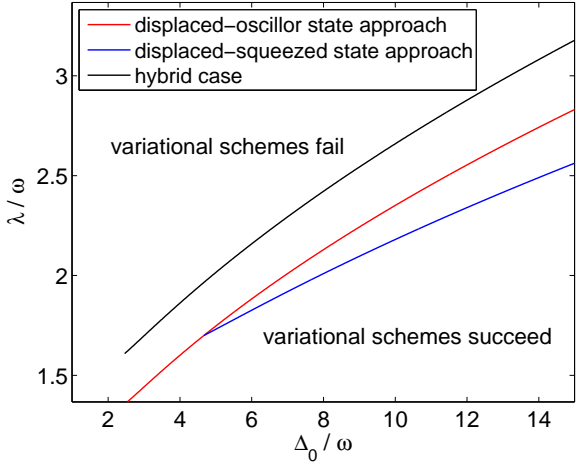}
\caption{(Color online) Fail point boundary lines of the three
variational schemes in the single-mode case. The starting points of
the curves are ($\Delta_0/\omega_c\simeq 2.54;$
$\lambda_c/\omega\simeq 1.37$), ($\Delta_0/\omega_c\simeq 4.67$;
$\lambda_c/\omega\simeq 1.70$), and ($\Delta_0/\omega_c\simeq 2.46$;
$\lambda_c/\omega\simeq 1.61$) respectively for the DOS, DSS, and
hybrid case.}
\end{figure}

It should be noted that the above analysis cannot be generalized
directly to the case of $T\not=0$. This is because the system with a
Hamiltonian given in Eq.\ (30) is not a true thermodynamical system.
To perform the variational calculation as in the multi-mode case,
one needs to include the environment as a ``thermo-bath", i.e., the
effect of the environment is just to keep the single-mode system in
thermodynamical equilibrium. Under this sense, we can do a
variational calculation based on the DOS and the up-bound of the
free energy in the present case is given by
\begin{equation}
f_a=-T\ln[2\cosh(\beta k_2\Delta_0/2)]-\omega^{-1}(C^2-2C\lambda),
\end{equation}
where
\begin{equation}
C=\frac{\lambda}{1+k_2\Delta_0\omega^{-1}\coth(\beta\omega/2)\tanh(\beta
k_2\Delta_0/2)},
\end{equation}
and $k_2$ is determined by the self-consistent equation
\begin{equation}
\ln k_2=-\frac{2\lambda^2\coth(\beta\omega/2)}{[\omega+k_2\Delta_0\coth(\beta\omega/2)\tanh(\beta k_2\Delta_0/2)]^2}.%\}.
\end{equation}
Correspondingly, the up-bound of the free energy of the
displaced-oscillator ground state is
\begin{equation}
f_0=-T\ln[2\cosh(\beta k_3\Delta_0/2)]-\lambda^2/\omega,
\end{equation}
where $k_3=\exp\{-2(\lambda/\omega)^2\coth(\beta\omega/2)\}$. Using
the above result, one can make analysis on the stability of
variational thermodynamical ground state as before. As expected, it
turns out that the situation is almost the same as in the case of
$T=0$. There is a frequency threshold $\omega_c$. For
$\omega>\omega_c$, we have $f_a<f_0$, i.e., variational
thermodynamical ground state is always stable for all the coupling
range and $k_2\Delta_0$ decreases continuously with increasing
$\lambda$; however, for $\omega<\omega_c$, variational
thermodynamical ground state becomes unstable as the coupling
reaches some critical value $\lambda_c$ which is again
$\omega$-dependent.
Tunneling splitting shows a discontinuous jump when the fail point is reached. %Fig.7 gives an illusion of free energy comparison and
%the discontinuous jump around the fail point.
In the case of $T\not=0$, the fail point can be reached in another
way; that is, for a given coupling $\lambda$, one can find that the
variational scheme fails when temperature rises to some critical
value $T_c$, at where tunneling splitting also shows a discontinuous
jump. The boundary line of the fail point for the variational scheme
based on the DOS in the case of $T\not=0$ is shown in Fig.\ 7. The
boundary lowers down as the temperature increases, which implies
that the variational scheme fails at a lower $\lambda_c$ as $T$
increases.
\begin{figure}[h]
\centering
\includegraphics[width=0.48\textwidth]{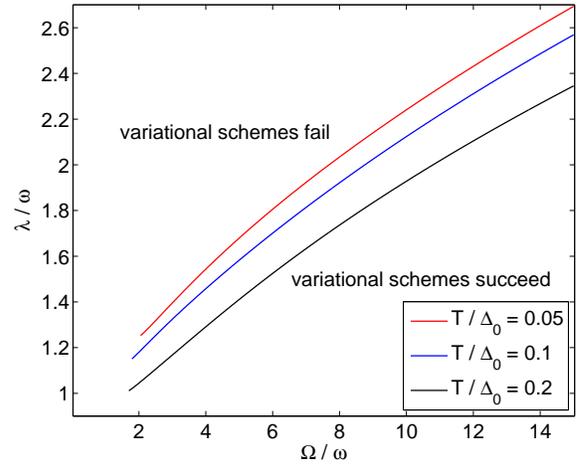}
\caption{(Color online) Boundary lines of fail point for
variational scheme based on the DOS in the single-mode case at
$T\not=0$. As $T$ raises, the boundary lines lowers down, showing
that the variational scheme fails at a lower $\lambda_c$ as $T$
increases. }
\end{figure}

Why variational schemes fail in the low frequency region?
Physically, in the high frequency region with $\omega/\Delta_0\gg
1$, the first term in Eq.\ (1) or (30) can be treated as a small
perturbation, hence the DOS is a good trial ground state for the
variational calculation; however, in the low frequency region with
$\Delta_0/\omega\gg1$, the DOS is not appropriate to serve as the
trial ground state since the term
$\frac{1}{2}\Delta_0\hat{\sigma}_x$ makes the main contribution in
the single mode case. As one can see from the analysis presented in
Ref.\ 4,  in the low frequency region, the true ground state of the
phonon bath in the single mode case is a combination of the two DOS
with opposite displacement, i.e.,
\begin{equation}
|\phi_{ph}\rangle=\frac{1}{\sqrt{1+a^2}}[e^{-g_1(\hat{b}^{\dagger}-\hat{b})}+a e^{g_2(\hat{b}^{\dagger}-\hat{b})}]|0\rangle,
\end{equation}
where $g_{1,2}$ and $a$ are functions of $\lambda/\Delta_0$ and
$\omega/\Delta_0$. Numerical analysis shows that
$|\phi_{ph}\rangle$ gives a very accurate description for the
phonon ground state in the low frequency region with strong
couplings. Obviously, $|\phi_{ph}\rangle$ can be described by
neither a DOS-like nor a DSS-like state, showing that both the DOS
and DSS can not serve as the trial ground states in the low
frequency region with strong couplings. As shown in Ref.\ 4, a
variational scheme by taking a trial ground state as
$|\phi_{ph}\rangle$ shows no discontinuous jump. We have made
numerical analysis on the energy of the variational ground state
based on $|\phi_{ph}\rangle$, which is found to be always stable
against the displaced-oscillator ground state in all the coupling
regime. This result clearly shows that the failure of the
variational scheme is due to the invalid trial ground state in the
low frequency regions. The above analysis provides the answer to
why the variational schemes based on both the DOS and DSS always
fail in low frequency region with strong couplings.

It should be noted that the low frequency modes play important roles
in the localized transition. As one can see from the self-consistent
equations (12) or (27), the integral can be divided into two parts
like, $\int_0^{\omega_D}\rightarrow
\int_0^{\omega_c}+\int_{\omega_c}^{\omega_D}$, without the
contribution of the low frequency modes lie in $(0,\omega_c)$, there
is no localization for any finite $\alpha$. In other words, it is
the low frequency modes lead to localization. The failure of the
variational schemes in the low frequency regions implies that the
variational calculations fail to take account of the contribution of
the low frequency modes to the localization, making the predicted
localization in question.

\subsection{Continuous or discontinuous crossover}
To the first order approximation (i.e., by omitting the cooperative
effect between different phonon modes), the result of multi-mode
case can be considered as a combination of all the phonon modes lie
in $(0,\omega_D)$ with a frequency-dependent weight. Generally
speaking, the parameter $s$ in the spectral function $J(\omega)$
determines the weight, while for counting the contribution of the
phonon modes that have fail point; the parameter $\Delta_0/\omega_D$
is therefore important and the boundary $\omega_c$ is in scale of
$\Delta_0$ as shown in Fig.\ 6. Let us take the variational scheme
based on the DOS as an example. Giving that $\Delta_0/\omega_D=0.1$,
one can see from Fig.\ 6 that $\omega_c/\Delta_0\simeq 1/3$, which
implies that the frequency modes with frequency
$0<\omega<0.03\omega_D$ have fail points, i.e., about 3 percent
phonon modes have fail points. However, if one takes
$\Delta_0/\omega_D=1$ as shown in Fig.\ 1, then there are about 30
percent phonon modes have fail points. This result shows that the
population of the phonon modes having fail points is directly
controlled by the value of $\Delta_0/\omega_D$. Now we know that
only the phonon modes have fail points will contribute a
discontinuous jump of the tunneling splitting as the coupling
reached the critical point. In the Ohmic case with
$\Delta_0/\omega_D\ll 1$, the discontinuous contribution of the low
frequency modes at the fail point is overwhelmed by the continuous
contribution from the high frequency modes, leading to a continuous
transition; however, as $\Delta_0/\omega_D$ increases to some larger
value (say, 0.5 in the DOS case), the increasing population of
phonon modes having fail points leads to an observable discontinuous
contribution and the transition behavior turns to discontinuous. In
the sub-Ohmic case, the weight of the low frequency modes increases
so that the discontinuous contribution from phonon modes having fail
points becomes observable even when $\Delta_0/\omega_D\ll 1$, hence
the transition is always discontinuous. However, for the same
reason, one can easily predict that the discontinuous transitions
will be unapparent when $\Delta_0/\omega_D$ becomes too small. It
should be noted that the discontinuous transition in the sub-Ohmic
case is just due to the fail point of the variational schemes, not
the true crossover behavior of the model since it is not expected to
be seen in the NRG or DMRG calculation.\cite{bulla,b1,wh1}

The above analysis can be generalized to the case of $T\not=0$. It
has been shown that the transition behavior predicted by the
variational calculations at $T\not=0$ is discontinuous for
$0<s<2$.\cite{ch} The key point is to understand why the
discontinuous transition behavior is extended to the super-Ohmic
case. By comparing the two self-consistent equations (12) for $T=0$
and (27) for $T\not=0$, at $T\not=0$, the spectral function is
effectively modified as $J(\omega)\coth(\beta\omega/2)$, i.e., an
extra factor $\coth(\beta\omega/2)$ appears in numerator of the
integrand. It is obvious that, $\coth(\beta\omega/2)\propto 1/\omega
$ as $\omega\rightarrow 0$, which implies that, in the low frequency
regions, $J(\omega)\coth(\beta\omega/2)\propto \omega^{s-1}$. In
other words, the extra factor $\coth(\beta\omega/2)$ seriously
increases the weight of the low frequency modes having fail points,
resulting in a discontinuous transition behavior even in the
super-Ohmic case.

Mathematical analysis on the self-consistent equations derived from
the variational condition at both $T=0$ and $T\not=0$ shows there
exists a universal $s_c$ for all the variational schemes, the
predicted transition is always discontinuous for $0<s<s_c$, a
continuous transition can only be found when $s=s_c$ with
$\Delta_0/\omega_D\ll 1$. It was found that $s_c=1$ for $T=0$ and
$s_c=2$ for $T\not=0$.\cite{ch} The above analysis clearly shows
that the predicted transition behavior depends on both the weight
and the population of the low frequency modes.

\section{Conclusion and discussion}
In conclusion, we have shown that all the variational schemes used
in the spin-boson model have fail points. As the fail point is
reached, the variational schemes fail to give a lower energy (or
lower free energy at $T\not=0$) than the displaced-oscillator ground
state, leading to a transition from the variational ground state to
the displaced-oscillator ground state, which is always misidentified
as the crossover from the delocalized to localized phases in a
variational calculation. Our analysis show that the existed
variational schemes are not suitable for studying the crossover
problem in the spin-boson model. The present analysis can help to
understand the controversy between different variational schemes and
discrepancy with other calculations, like the NRG and DMRG. The
analysis on the fail point in the single mode case can help to
answer why the variational schemes have fail points and how
different transition behaviors were found in different conditions.
However, the situation is still far from totally satisfactory. %In fact, it can be
%shown that, in both variational schemes, the transition point to a
%state with zero tunneling splitting is always a fail point of the
%variational scheme, showing that the predicted crossover point is
%not reliable. However,
One intriguing fact is that the ``fake" crossover point predicted by
the variational calculation based on the DOS is found to be in good
agreement with the results obtained by other methods including the
NRG calculation for both the Ohmic and sub-Ohmic
cases.\cite{leg,bulla,wh} In the present stage, it is not yet clear
if this is just a coincidence. One possible explanation is that
since the population of the low frequency modes having fail points
is very small in the case of $\Delta_0/\omega_D\ll 1$, the
contribution of them is small comparing with the rest of the phonon
bath. Our DMRG calculation also assures that the variational ground
state based on the DOS is a good approximation to the true ground
state before the transition happens.\cite{wh1} Nevertheless, as we
have mentioned before, the phonon-induced localization is mainly due
to the low frequency modes, combined with the discontinuous
transition behavior in the sub-Ohmic case, the answer is not very
convincing. Perhaps one can find that answer by developing a
variational scheme without fail point and this work is now under
process. The present result can also shed some lights on the failure
of the variational calculation on the self-trapped transition in
ploaron and exciton systems. In fact, the present analysis is
readily applicable to the discontinuous transition predicted by the
variational calculation gave in Ref.\ 4, but further analyses are
needed for other cases.\cite{la,gl}
%In adiabatic approximation, it is the low frequency modes leading to the localization,
%while variational schemes fail to take account of the contribution of the low frequency modes,
%which makes the result from variational calculations unreliable.
\begin{acknowledgements}
This work was supported by a grant from the Natural Science
Foundation of China  under Grant No. 10575045.
\end{acknowledgements}


\begin{thebibliography}{99}
\bibitem {note}Using the notation of Ref.\ \onlinecite{cy}, the result is $\alpha_c=3-2\lambda$. From Eq.\ (11) in
this reference, it is found that $s=2\lambda+n-1$, for $s=1$, we
have $2\lambda=2-n$, this leads to $3-2\lambda=n+1$.
\end{thebibliography}
\end{document}